\documentstyle[12pt]{article}

\def\be{\begin{equation}}
\def\ee{\end{equation}}
\def\bea{\begin{eqnarray}}
\def\eea{\end{eqnarray}}

\newcommand{\bean}{\begin{eqnarray*}}
\newcommand{\eean}{\end{eqnarray*}}

\def\br{\begin{thebibliography}{abc}}
\def\er{\end{thebibliography}}
\def\rf{\bibitem}

\def\cstar{C$^*$-algebra }

\newcommand{\I}{\mbox{\rm I} \hspace{-0.5em} \mbox{\rm I}\,}

 \newcommand{\complex}{
        \mbox{C \hspace{-1.16em} \raisebox{-0.018em}{\sf l}}\;}
\def\iff{\Leftrightarrow}

\def\bar#1{\overline{#1}}

\def\Hat#1{\rlap{\kern.10em$\widehat{\phantom G}$}#1}
\def\HAt#1{\rlap{\kern.05em$\widehat{\phantom G}$}#1}

\def\czp#1{\rlap{\kern.1em$\widehat{\phantom{G\vrule height.8em}}$}#1{}}
\def\Czp#1{\rlap{\kern.05em$\widehat{\phantom{G\vrule height.8em}}$}#1{}}

\newcommand{\sect}[1]{\setcounter{equation}{0}\section{#1}}

\newcommand{\nn}{{^n}}
\newcommand{\inff}{{^\infty}}

\newcommand{\A}{\mbox{${\cal A}$}}
\newcommand{\An}{\mbox{${\cal A}_{n}$}}
\newcommand{\Ainf}{\mbox{${\cal A}_{\infty}$}}

\def\ca{{\cal A}}
\def\ch{{\cal H}}
\def\alg{{\cal A}_{\infty}}
\def\q{Q^{\infty}}
\renewcommand{\thefootnote}{\fnsymbol{footnote}}
\def\fn{\footnote}
\def\sxn#1{\bigskip\medskip \sect{#1} \smallskip
                                                 }

\begin{document}

\thispagestyle{empty}
\setcounter{page}{0}

\begin{flushright}

ESI 239 (1995)\\
DSF-T-35/95\\
UICHEP-TH/95-6\\
hep-th/9507147\\

\hfill July 1995
\end{flushright}

\vspace{.25cm}

\centerline {\Large  LATTICES AND THEIR CONTINUUM LIMITS}
\vspace{1cm}
\centerline {\large G. Bimonte$^{1,}$\fn{Address after September
$1^{\rm st}: $ Departamento de Fisica Teorica, Facultade de Ciencias,
Universitad de Zaragoza, 50009 Zaragoza, Spain.},
                    E. Ercolessi$^2 $,
                    G. Landi$^{3,4}$,}
\vspace{2mm}
\centerline{\large  F. Lizzi$^{4,5}$,
                    G. Sparano$^{4,5}$,
                    P. Teotonio-Sobrinho$^6$}
\vspace{1cm}
\centerline {\it The E. Schr\"odinger International Institute for
Mathematical Physics,}
\centerline{\it Pasteurgasse 6/7, A-1090 Wien, Austria.}
\vspace{2mm}
\centerline {\it $^1$ International Centre for Theoretical Physics,
P.O. Box 586, I-34100, Trieste, Italy.}
\vspace{2mm}
\centerline {\it $^2$ Dipartimento di Fisica and INFM, Universit\`a di
Bologna,}
\centerline{\it Via Irnerio 46, I-40126, Bologna, Italy.}
\vspace{2mm}
\centerline{\it $^3$ Dipartimento di Scienze Matematiche,
Universit\`a di Trieste,}
\centerline{\it P.le Europa 1, I-34127, Trieste, Italy.}
\vspace{2mm}
\centerline {\it $^4$ INFN, Sezione di Napoli, Napoli, Italy.}
\vspace{2mm}
\centerline {\it $^5$ Dipartimento di Scienze Fisiche, Universit\`a di
Napoli,}
\centerline{\it Mostra d' Oltremare, Pad. 19, I-80125, Napoli, Italy.}
\vspace{2mm}
\centerline {\it $^6$ Dept.\ of Physics, Univ.\ of  Illinois at Chicago,}
\centerline{\it 60607-7059, Chicago, IL, USA.}
\vspace{.2cm}
\begin{abstract}
We address the problem of the continuum limit for a system of
Hausdorff lattices (namely lattices of isolated points)
approximating a topological space $M$.
The correct framework is that of projective
systems. The projective limit is a universal space from which $M$ can
be recovered as a quotient. We dualize the construction to
approximate the algebra ${\cal C}(M)$ of continuous functions on $M$.
In a companion paper
we shall extend this analysis to
systems of noncommutative lattices (non Hausdorff lattices).
\end{abstract}

\newpage
\setcounter{page}{1}

\renewcommand{\thefootnote}{\arabic{footnote}}
\setcounter{footnote}{0}

\sxn{Introduction}\label{se:1}

Lattice discretizations have become very popular methods to
approximate physical models which are too complicated to be solved
analytically \cite{lattice}.
However, in spite of their success, there are certain features of
continuum dynamics which are generally not addressed in a transparent
and satisfactory way.
For example, it is not obvious how to describe any topological aspect of
quantum physics within a lattice approach.

A typical continuum theory is usually given by a suitable carrier
space (configuration or phase space) together with a dynamics on it.
Interesting properties of the physical system could come from either
of them. For instance, nontrivial topological properties of the
configuration space may have deep consequences even for simple
dynamics. On the contrary, in the usual lattice models, these two
aspects are not clearly separated.

It is interesting to formulate lattice theories in a way that dynamical and
kinematical aspects remain as separated as possible. The first question one can
ask is how the topology of the underlying space (-time) $M$ arises from a
lattice of points, regardless of the specific dynamics. The second, and more
difficult question, refers to the topology of the ($\infty $-dimensional) space
$\Gamma $ of all configurations. In typical lattice models, the only
topological information refers to $M$ and is that of nearest neighbors as
encoded in the Hamiltonian. Even though this captures some of the global
topological features of $M$ it does not provide {\em per se} a notion of limit
in which $M$ is recovered. Moreover, this incomplete topological information
has no bearings on the configuration space $\Gamma $ which is topologically
trivial. For instance, this is the reason why on the lattice solitons are not
truly topological.

In \cite{BBET,NCL} we have initiated a systematic investigation
of these issues. This work has been inspired by a paper of Sorkin
\cite{So}, where it is shown how a Hausdorff topological space can be
approximated with finite, non Hausdorff topological spaces (posets).
This method gives satisfactory results under two aspects. On one
side, already with a finite number of points it reproduces relevant
topological properties of the space being approximated. On the other
side, it gives, via the notion of projective system, a well defined
concept of continuum limit from which the initial space can be
reconstructed.
In \cite{BBET,NCL} we developed the essential tools
for doing quantum physics on finite topological spaces and considered
the dualization of these spaces.
In \cite{NCL}
it was observed that
posets are genuine noncommutative spaces in the sense that
one can associate with them a noncommutative algebra
${\cal A}$ playing a role analogous to
that of the algebra of continuous functions for Hausdorff spaces\fn{These
algebras contain
enough information to reconstruct the lattice completely, thus
providing a full dualization.}.
This algebraic framework provides new and well developed tools to
construct quantum mechanical and field theoretical models.
Connes' noncommutative geometry \cite{Co} for example, can be
immediately applied giving access to structures that retain their
richness even when the geometry, though not trivial, is anyway poorer
than the one of the
continuum\footnote{
For example in \cite{BLS} the noncommutative geometry of the distances is
applied to lattices.}.
In this way, topological
informations enter non trivially at all stages of the construction.
In that paper we have also explicitly shown how non trivial
topological effects are captured by these topological lattices and
their algebras, by constructing algebraically the $\theta$-quantizations
of a particle on a circle poset.

In this and a companion paper \cite{comp}, we address the question of how to
recover a topological space $M$ through a suitable notion of limit of  a system
of lattices $Q^n$. The dualization of this framework, in the spirit of
noncommutative geometry, is also analyzed. We show how the algebra ${\cal
C}(M)$ of continuous functions on $M$ can analogously be recovered from the
system of algebras ${\cal C}(Q^n)$.

In the present paper, we apply these methods to Hausdorff lattices,
namely to lattices made of isolated points. This kind of lattices
arises, for example,
when discretizing a scalar field defined on a manifold $M$.
We shall show that, even in this simple case,
a structure of projective system \cite{EDM}
produces a topologically non trivial limit space from which a
topological space
$M$ being approximated can be recovered.
As we shall see, the limit space is a universal one in the sense that
any two such limit spaces for different spaces $M$
are naturally homeomorphic and can be
identified with the Cantor set.
The extra information which is needed to recover $M$ is provided by a
projection $\pi$ from the Cantor set to $M$.
This projection, which is not naturally built in the limit space,
can be constructed starting from the projective system.

We then show that there is a structure of direct (or inductive) limit
\cite{EDM} on the algebra of continuous functions defined on the
lattices. Although at a finite level these algebras are trivial,
their inductive limit is the algebra of continuous functions on the Cantor
set. The algebra ${\cal C}(M)$ of continuous functions on $M$ is then
the subalgebra of projectable functions with respect to $\pi$.

Since Hilbert spaces play a key role in quantum theories and also in
noncommutative geometry,
a similar analysis will be repeated for $L^2(M)$
through an inductive system of finite dimensional Hilbert spaces.

As mentioned before, in \cite{comp} we shall see that with the use of
topological lattices the projective and the inductive limits will
loose they universal character and will be naturally related to $M$
and ${\cal C}(M)$ respectively.

\sxn{Continuum limit of Hausdorff Lattices}

Consider a piecewise linear space $M$ of dimension $d$, or in other words
a space that admits a locally finite cellular decomposition
$\Sigma = \{S_\alpha, \alpha\in I, I\subset {\bf N}\}$. For
convenience we shall use cubic cells so that the
$S_\alpha$'s will be closed cubes. Once such a
decomposition is chosen, one can associate with it a lattice $Q$ of
points. The way this is usually done is by looking at the set of
vertices in the decomposition. In this paper, however, we will
instead
take the vertices of the dual lattice, which means that the points of
$Q$ correspond to highest dimensional cubes. With this choice it
becomes possible to introduce a nontrivial notion of limit for a
sequence of finer and finer decompositions.

The lattice $Q$ is then given a Hausdorff topology. On a space with a
finite (or countable) number of points there is a unique Hausdorff
topology, and it is the one for which each point is open and closed at
the same time.
Our aim is to understand if, and to what extent, $M$ can be recovered
as a limit.
We then consider
a sequence $\Sigma^n$ of finer and finer cubic decompositions,
for example the one obtained by splitting
every $d$-dimensional cube in two at each step, together with
their associated lattices $Q^n$'s.

Now, the simple fact of having a sequence of lattices is not
sufficient yet to obtain $M$ as its limit. What one has to do is to give
this sequence a further structure which converts it into what is
known in mathematics as an inverse or projective system of topological spaces,
which we now pass to describe.

A  {\em projective} (or {\em inverse}) {\em system}  of  topological spaces
is a family of topological spaces $Y\nn , n \in {\bf N}$ \footnote{More
generally, the index $n$ could be taken in any directed set. } together with a
family of continuous projections
$\pi^{(m,n)}:Y^{m}\to Y\nn,\ n\leq m$, with the
requirements that $\pi^{(n,n)}=\I,\ \pi^{(n,m)}=\pi^{(n,p)}\pi^{(p,m)}$.
The projective limit
$Y^\infty$ is defined as the set of coherent sequences, that is the set of
sequences $\{x\nn\in Y\nn\}$ with
$x\nn=\pi^{(m,n)}(x^{m})$.

There is a natural projection
$\pi^{n} : Y^\infty \rightarrow Y\nn $ defined as:
\be
\pi^{n}(\{x^{m}\in Y^{m} \})= x^{n}~.
\ee
The space $Y^{\infty}$ is given a topology, by declaring that a set
${\cal O}^{\infty}\subset Y^{\infty}$ is open iff it is the inverse image of
an open set belonging to some $Y^{n}$ or a union (finite or
infinite) of such sets.

Let us then consider again the sequence of cubic decompositions
$\Sigma\nn = \{S^n_\alpha, \alpha\in I^n\}$, with
$\Sigma^{(n+1)}$ obtained from $\Sigma\nn$ by subdivision of its cubes.
In order
to be able to correctly reproduce the space $M$ in the limit, the sequence
of cubic decompositions
must be such that all cubes in it become smaller and smaller. The
precise meaning of this requirement is that for any point $x\in M$
and any open set ${\cal O}_x$ containing $x$, there must exist a
level of approximation such that all cubes containing $x$ will be
contained in ${\cal O}_x$ from that level on\fn{We are really using
decompositions which are `fat' in the sense of \cite{CMS}.}:
\be
\forall~ x~ {\rm and}~ \ \forall~ {\cal O}_x \ni x,\ \exists~ m~
{\rm ~such~ that~}~ \forall~ n\geq m~,
S\nn_\alpha\ni x\  \Rightarrow
\ S\nn_\alpha\subset{\cal O}_x~. \label{reqseq}
\ee

This subdivision procedure
naturally induces a structure of projective system on the
corresponding sequence $Q^n$ of lattices.

The projection
\be
\pi^{(m,n)}:Q^{m}\to Q\nn,\ m>n   \label{pro}
\ee
associates to a $d$-dimensional cube of the finer subdivision the
unique
$d$-dimensional cube from which it comes.

\begin{figure}[htb]
\begin{center}
\unitlength=1.00mm
\special{em:linewidth 0.4pt}
\linethickness{0.4pt}
\begin{picture}(147.00,75.00)
\put(35.00,75.00){\makebox(0,0)[cc]{$\vdots $}}
\put(111.00,75.00){\makebox(0,0)[cc]{$\vdots $}}
\put(5.00,29.00){\line(0,1){4.00}}
\put(5.00,44.00){\line(0,1){4.00}}
\put(65.00,29.00){\line(0,1){4.00}}
\put(65.00,33.00){\line(0,1){0.00}}
\put(65.00,44.00){\line(0,1){4.00}}
\put(65.00,59.00){\line(0,1){4.00}}
\put(35.00,29.00){\line(0,1){4.00}}
\put(35.00,44.00){\line(0,1){4.00}}
\put(35.00,59.00){\line(0,1){4.00}}
\put(50.00,44.00){\line(0,1){4.00}}
\put(50.00,59.00){\line(0,1){4.00}}
\put(20.00,44.00){\line(0,1){4.00}}
\put(20.00,59.00){\line(0,1){4.00}}
\put(12.00,59.00){\line(0,1){4.00}}
\put(27.00,59.00){\line(0,1){4.00}}
\put(42.00,59.00){\line(0,1){4.00}}
\put(57.00,59.00){\line(0,1){4.00}}
\put(76.00,61.00){\circle*{2.00}}
\put(86.00,61.00){\circle*{2.00}}
\put(96.00,61.00){\circle*{2.00}}
\put(106.00,61.00){\circle*{2.00}}
\put(116.00,61.00){\circle*{2.00}}
\put(126.00,61.00){\circle*{2.00}}
\put(136.00,61.00){\circle*{2.00}}
\put(146.00,61.00){\circle*{2.00}}
\put(81.00,46.00){\circle*{2.00}}
\put(86.00,58.00){\vector(-1,-3){2.67}}
\put(76.00,58.00){\vector(1,-3){2.67}}
\put(101.00,46.00){\circle*{2.00}}
\put(106.00,58.00){\vector(-1,-3){2.67}}
\put(96.00,58.00){\vector(1,-3){2.67}}
\put(121.00,46.00){\circle*{2.00}}
\put(126.00,58.00){\vector(-1,-3){2.67}}
\put(116.00,58.00){\vector(1,-3){2.67}}
\put(141.00,46.00){\circle*{2.00}}
\put(146.00,58.00){\vector(-1,-3){2.67}}
\put(136.00,58.00){\vector(1,-3){2.67}}
\put(91.00,31.00){\circle*{2.00}}
\put(100.00,43.00){\vector(-3,-4){6.00}}
\put(82.00,43.00){\vector(3,-4){6.00}}
\put(131.00,31.00){\circle*{2.00}}
\put(140.00,43.00){\vector(-3,-4){6.00}}
\put(122.00,43.00){\vector(3,-4){6.00}}
\put(111.00,16.00){\circle*{2.00}}
\put(129.00,28.00){\vector(-3,-2){14.00}}
\put(93.00,28.00){\vector(3,-2){14.00}}
\put(35.00,5.00){\makebox(0,0)[cc]{(a)}}
\put(111.00,5.00){\makebox(0,0)[cc]{(b)}}
\put(5.00,16.00){\line(1,0){60.00}}
\put(65.00,31.00){\line(-1,0){60.00}}
\put(5.00,46.00){\line(1,0){60.00}}
\put(65.00,61.00){\line(-1,0){60.00}}
\put(5.00,59.00){\line(0,1){4.00}}
\put(65.00,14.00){\line(0,1){4.00}}
\put(5.00,14.00){\line(0,1){4.00}}
\end{picture}
\end{center}
\begin{center}
{\footnotesize {\bf Fig. 1.} (a) shows a subdivision of the interval.
(b) shows the corresponding projective system.}
\end{center}
\end{figure}

We call
$Q\inff(M)$ the projective limit of the projective system.
A point in $Q^\infty(M)$ is nothing
but a decreasing sequence $\{q\nn\}$ of cubes, namely a sequence such
that $q^{n+1}\subset q\nn$.
We shall see that this space is not the original space $M$, but
it is bigger in the sense that
$M$ can then be
recovered from it as a quotient.
For simplicity we will use $q^{n}$ to denote both the element of
$Q^{n}$ and the corresponding cube $S\nn_\alpha$. There exists a natural
projection $\pi : Q^{\infty }(M) \rightarrow M$. It is defined as
follows:
\be
\pi(\{q\nn\}) =   \bigcap_n  q\nn. \label{prob}
\ee

In this manner, we get a unique point of $M$. That this point is
unique is a consequence of condition (\ref{reqseq}).

In order to illustrate the topology of $Q^{\infty }(M)$ we will
consider the case when $M$ is the interval $I=[0,1]$. Its
decomposition and the associated projective system are shown in Fig.
1.

An element of $Q^{\infty }(I)$, namely a coherent sequence
$q^{\infty } \equiv \{ q^n \in Q^n \}$, can be identified with a
string $\epsilon _1 \epsilon _2 \epsilon _3 ... $
of $0$'s and $1$'s and the
correspondence is one to one. The string can be constructed in the
following way: The starting approximation  $Q^0$ consists of a
single element corresponding to the whole interval. After the first
subdivision the interval $I$ is split into two equal halves. Then we take
$\epsilon _1$ equal to $0$ or $1$ depending on whether $q^1$ is the
left or right half; $\epsilon _2$ will similarly be $0$ or $1$
depending on whether $q^2$ is the left or right half in which
$q^1$ splits in going from $Q^1$ to $Q^2$ and so on.
With the help of a decimal point on the extreme left, we can see this
sequence as the binary representation of a point on the interval.
Notice that this point coincides with $\pi(q^{\infty})$. This
labeling of $Q^\infty (I)$ makes it clear that there might be more
than one point in $Q^\infty (I)$ that project to the same point of
$I$.
For example the points $.0111\cdots$ and $.1000\cdots$ both correspond to the
point $1/2$.
The same thing will happen for all points of $I$ of the form
$m/2^n$, with $m$ and $n$ integers. On the other hand,
the remaining points
of $I$ are the image of a unique point of $Q^{\infty}(I)$,
because they have a unique binary representation.

Thus $Q^{\infty}(I)$
is a ``quasi fiber bundle" over $I$ whose fibers contain either one
or two points.
Even though $Q^{\infty}(I)$ is in this sense bigger than $I$, the
interval can be recovered as the quotient $Q^{\infty}(I)/\sim$ ,
where $\sim$ indicates the equivalence
relation defined by the projection $\pi$. From the topological
point of view,
this statement is nontrivial, as it is not guaranteed that
$Q^{\infty}(I)/\sim$ , endowed with the quotient topology, is
homeomorphic to $I$. The proof that this is the case will be given
below in the general case.

The space
$Q^ \infty (I)$ is nothing but a Cantor set which is, up to
homeomorphisms, the only totally disconnected
\footnote{ A totally disconnected space is one for which
each connected component is just a single point.},
perfect\footnote{A perfect space is one for which each of its points
is an accumulation point for the space.},
metric topological space.
A familiar realization of a Cantor set is the ``middle-third Cantor
set". It is obtained by starting from the interval $[0, 1]$
dividing it in three parts, removing the middle third and iterating
the procedure ad libitum on each of the remaining parts.
That the previous
$Q^{\infty}(I)$ is a Cantor set can be then proven either by showing
directly that it enjoys the mentioned properties or by explicitly
showing that it is homeomorphic to the middle third Cantor set.

The space  $Q^{\infty}(I)$ coming from the interval is actually
``universal", in the sense that it is homeomorphic to the space
$Q^{\infty}(M)$ associated with a generic $M$. A simple argument
to show that
this is the case goes as follows: a projective system for $M$ can
always be obtained by taking some initial decomposition of $M$ in
cubes and
by refining it by suitably splitting each cube
in two halves at each step. It is then clear that the corresponding
projective system coincides with the one we have constructed for
the interval from a certain interval on, and thus has the same
projective limit. The difference lies in the projection $\pi$, whose
definition, as given in equation (\ref{prob}), uses the explicit
interpretation of the
$q\nn$ as subset of $M$ and thus depends on
the specific $M$. Again $Q^{\infty}(M)$ is a quasi fiber bundle over
$M$. The number of points in the fibers also depends on the
projection $\pi$. In other words:
{\em  $Q^{\infty}(M)$ is universal, but the
fibration is not.}
In fact, the existence of such a projection $\pi$ is
not surprising due to a known result \cite{HY} that there exists a
continuous projection from the Cantor set onto any compact metric
topological space.
For this reason, from now on, we simply write $Q^\infty$ instead of
$Q^\infty(M)$.

We now turn to the proof that $M$ is actually homeomorphic to the
quotient space $Q^\infty /\sim$ .

Let us first prove that the projection $\pi$ in eq.(\ref{prob}) is
continuous.
We have to  show that the inverse image of an open set $B$ in $M$ is
open in $Q^\infty(M)$. Let $q= \{S^n_{\alpha}(q)\}$ be a point
belonging to $\pi^{-1}(B)$ and let $x = \pi(q)$. Because of the
condition (\ref{reqseq}) on the sequence of cubical decompositions there exists
a
$j\in N$ such that $n > j$ implies that all cubes
$S^n_{\alpha}$ containing $x$ are all contained in $B$.
Consider then ${\cal O}^\infty =
\pi^{-1}(S^n_{\alpha}(q))$ with $n>j$ which is an open
set of $P^\infty(M))$ containing $q$. ${\cal O}^\infty$ is also entirely
contained in $\pi^{-1}(B)$, in fact all its points are coherent
sequences  whose representatives at level $n$ are cells fully
contained in $S^n_{\alpha}(q)$ and since the cube
$S^n_{\alpha}(q)$ is fully contained in $B$ they project in $B$.

To prove that the topology of $M$ is equivalent to the quotient topology
on $Q\inff /\sim$ , it is then sufficient to show that
the inverse image of a subset  of $M$, which is
{\em not}  open, is not open in $Q\inff$ as well.

Consider then the set $\pi^{-1}(B)\subset  Q\inff$, with
$B \subset M$ {\em not} open. We will show that the assumption
that $\pi^{-1}(B)$ is open leads to a contradiction.

The statement that $B$ is not open in the topology of $M$ is
equivalent to saying that there exists a sequence of points
$\{ x_i \}$ of $M$, not belonging to $B$, which converges to a point
$x \in B$. From this sequence we will extract a particular subsequence
$\{y_j \}$, still converging to $x$. We first introduce a countable basis
of open neighborhoods for $x$, namely a countable
family $\{ {\cal O}_i\}$ of decreasing open sets containing $x$.

Let us start with $ {\cal O}_1$.
Due to condition (\ref{reqseq}), there are one or more $d$-cubes
$S_\alpha^{n(1)} \subset { \cal O}_1$, with $ S_\alpha^{n(1)} \ni x$.
At least one of these $d$ cubes,
call it $S_{\alpha(1)}^{n(1)}$, will contain an infinite number of elements
the sequence $\{x_i\}$.
Then, choose $y_1$ to be any one of these elements.

At the next level $2$ there will again be at least one $d$-cube
$  S_{\alpha(2)}^{n(2)} \subset S_{\alpha(1)}^{n(1)} $,
with $ S_{\alpha(2)}^{n(2)}$ still containing an
infinite number of elements of the sequence ${x_i}$.
Again choose $y_2$ as any one of these elements.

By iterating this procedure, we obtain the sequence
$\{y_j \}$, which, being extracted from the
original sequence, still converges to $x$.
Moreover, $y_j \in S_{\alpha(j)}^{n(j)} $
and $\{ S_{\alpha(j)}^{n(j)} \}$ is a coherent sequence
\footnote{A little care must be taken as $n(j)$ may not coincide
with the subdivision level of the lattice.}
which thus defines a point $q\in Q \inff$. By construction
$\bigcap_j S_{\alpha(j)}^{n(j)} = x$, and consequently $\pi(q)=x$.

Since $\pi^{-1}(B)$ is assumed to be open, and
recalling how the topology of $Q\inff$ is defined,
there will be a $\bar{j}$ such that $\pi^{-1}(S_{\alpha(j)}^{n(j)})
\subset \pi^{-1}(B)$ for $j \geq \bar{j}$.
But then also
$\pi^{-1}(y_j)$, with $ j \geq \bar{j}$ , must belong to
$\pi^{-1}(B)$ and this implies,
contrary to the hypothesis on the sequence $\{x_i\}$, that $y_j\in B$.

\sxn{Algebras for Hausdorff Lattices}

In the previous Section, we have shown how to approximate a
topological space $M$ by a sequence of lattices and how to recover
$M$ by a limiting procedure. Here we shall dualize this construction.
The projective system of lattices will be replaced by an
inductive system of commutative $C^*$-algebras, and $Q^{\infty}(M)$
by the inductive limit ${\cal A}_{\infty}$. While before $M$ could be
recovered as a quotient of $\q(M)$, now ${\cal C}(M)$ will turn out to be a
subalgebra of $\alg$.
Duality here is understood in the sense of the Gel'fand-Naimark
theorem \cite{FD}.
The idea behind this theorem is that the full topological
information on a Hausdorff topological space $M$ is encoded in the
abelian $C^*$-algebra ${\cal C}(M)$ of its continuous functions. The space
itself is identified with the set of all complex homomorphisms of
${\cal C}(M)$ and the topology is given in terms of pointwise convergence:
\be
p_n \rightarrow p~~~~~\iff~~~~f(p_n)\rightarrow f(p)~~~\forall f
\in {\cal C}(M)~,~~~p_n,~p \in Hom({\cal C}(M),{\complex })~.
\ee

To each $Q\nn$ we now associate its algebra $\An$
of continuous functions. Since the $Q\nn$'s are discrete Hausdorff
spaces, a continuous function ${a_n}$ is just an assignment of a complex
number to each point $q^n$ in $Q\nn$, and then
\footnote{ Again we assume that the
lattices split in half going from one level to the next, so that the total
number of points at level $n$ is $2^n$.}
${\cal A}_n \equiv {\complex }^{2^n}$.
We shall write any such a function as the vector
\be
a_n = \{\lambda_1, \cdots, \lambda _{2^n}\}~.
\ee
The norm $||~ \cdot~ ||_n$ of a function $a_n$ is just the sup norm.

While in the previous
Section the framework for defining a limiting procedure was that of a
projective system of topological spaces, here it will be that of a
direct or inductive system of $C^*$-algebras.

An {\it inductive system} of $C^*$-algebras is a sequence of $C^*$-algebras
$\An$, together with norm non-increasing immersions
$\Phi_{(n,m)}~:~\An \rightarrow \A_m,~n<m$, such that the
composition law $\Phi_{{(n,m)}}\Phi_{{(m,p)}}=\Phi_{{(m,p)}},~n<m<p~,$ holds.

The inductive limit $\ca_{\infty}$
is the \cstar consisting of equivalence classes
of ``Cauchy sequences" $\{a_n\}, \ \ a_n\in
{\cal A}_n$ . Here by Cauchy
sequence we mean that $||\Phi_{{(n,m)}}(a_n) -a_m||_m$ goes to zero
as $n$ and $m$ go to infinity.
Two sequences $\{a_n\}$ and $\{b_n\}$ are equivalent if
$||a_n - b_n||_{n}$ goes to zero. The norm in $\ca_\infty$ is defined by
\be
||a||_\infty = \lim_{n\rightarrow\infty} ||a_n||_{n}
\ee
where $\{a_n\}$ is any of the representatives of
$a$ \fn{For a more
detailed account of the definition see for example \cite{FD} or
\cite{Mu}.}.

In our case the direct system is naturally defined by the pull-backs
$\Phi_{{(n,m)}}=\pi^{(n,m)*}$ associated with the
projections  in (\ref{pro})
\be
(\Phi_{(n, m)}(a_n))(q_m) = a_n (\pi^{(m, n)}(q_m)) \ . \label{emb}
\ee
The $\Phi$'s are isometric $*$-homomorphisms. Where previously there was
a projection $\pi^{(n)}$ from $\q$ to $Q\nn$, there is now an immersion
$\Phi_n$ of $\An$ in $\Ainf$ defined as:
\be
\Phi_n(a_n)=\{\Phi_{{(n,m)}}(a_n),~n<m  \},~a_n\in \A_n~.
\ee

The algebra $\Ainf$ is isomorphic to the \cstar of continuous functions on
$\q$. In order to prove this, it is useful to realize $\q$ as the
middle third Cantor set introduced earlier. Now an element $a_n
\in \An$ identifies an element $\Phi_n(a_n)\in \Ainf$ that can
be thought of as
a continuous and piecewise constant function on $Q\inff$.
The collection of all such functions is dense in $\alg$ by the very
definition of $\Ainf$.
Therefore every element of $\Ainf$ can be thought of as a uniformly
convergent sequence of functions of this type, and naturally determines a
continuous function on $Q\inff$. Conversely, we can now prove that
for any continuous
function $f$ on $\q$, one can find a sequence
$\{f_n\}$ of functions in $\Ainf$ uniformly
converging to it.
Indeed, since the Cantor set is a compact metric space, every
continuous function is also uniformly continuous, thus
$\forall \epsilon > 0~ \exists ~\delta_n$ s.t. for $| x-x' | <
\delta_n~,~| f(x) - f(x') | < {1\over 2^n}$.
The sequence $\{ f_n\}$ is defined
as follows: $f_n$ is a continuous piecewise constant function on $Q\inff$
defined by an element $\bar{f}_n\in {\cal A}_{m(n)}$, with $m(n)$ such that
${1\over 3^{m(n)}}\leq \delta_n$. The value of $f_n$ in any of the sets in
which it is constant is simply any of the values of $f$ in that set.

We now have to recognize in $\ca_{\infty}$
the algebra ${\cal C}(M)$ of continuous
functions on the space $M$. We will show
that ${\cal C}(M)$ is a subalgebra of \Ainf, which is the dual
statement of the fact that $M$ is a quotient of $Q\inff$.

In this respect we remind that $Q\inff$ is a quasi fiber bundle on
$M$. We will now show that the
algebra ${\cal C}(M)$ of continuous functions on $M$ is isomorphic to the
subalgebra of $\alg$ made of projectable functions. To start with, it is
obvious that the pull-back of continuous functions on $M$ are continuous
functions on $\alg$, which take constant value on the fibers. It is
sufficient to show then that they exhaust all such functions of $\alg$.
Consider then a continuous function, $f \in \alg$ which is constant on the
fibers. The function
$f$ thus defines naturally a function $\tilde f$ on $M$. The inverse
image ${\cal O}^{\infty}=f^{-1}({\cal O})$
of an open set, $\cal O$ of $\complex$  is an open set
of $Q\inff$ containing all the fibers through its
points. Since we have already shown that $M$ is homeomorphic
to $Q/\sim$, then
${\cal O}^\infty$ projects onto an open set, ${\cal O}_M$, of $M$. Since
${\cal O}_M$ is the inverse image of $\cal O$ through $\tilde f$,
$\tilde f$ is itself continuous.

Seen from the base, a generic element of $\alg$ can be regarded as a
multi-valued function on $M$ while ${\cal C}(M)$ can be identified with the
set of projectable functions on $\q$.

We thus have proven that the algebra of continuous functions over the Cantor
set $Q\inff$ contains as subalgebra the algebras of all continuous functions
over
compact topological spaces.

\sxn{Hilbert Spaces}

Since Hilbert spaces and representation of algebras of observables as
operator play a prominent role in quantum mechanics, we now show how
these structures fit in our scheme. In particular we will see how the
space $L^2(M, \mu )$ of square integrable functions of $M$ can be
approximated by the analogous spaces $L^2(Q^n, \mu^n)$ for the
lattices $Q^n$ and recovered as an inductive limit.

Now, the algebras $\An$ have an obvious representation as
diagonal matrices on the Hilbert space $\ch ^\nn = \complex ^{2^n}$,
given by
\be
a_n = (\lambda_1, ..., \lambda_{2^n}) \in \ca_n~\rightarrow
\hat{a}_n = \mbox{diag}(\lambda_1, ...,
\lambda_{2^n})\in M_{2^n}({\complex }).
\ee
The inner product between two vectors $\phi=(\phi _1, ...,\phi _{2^n})$
and $\psi=(\psi _1, ...,\psi _{2^n})$ of $\ch^n$ is defined as
\be
<\psi_n , \phi_n > = \sum _{i=1}^{2^n}\psi ^*_i\phi _i \mu_i^n
\ee
where $\mu _i^n$ are the normalized measures of the cubes
associated with the points of $Q^{(n)}$.

What we have is a sequence $\ch ^n$ of Hilbert spaces, one for
each level. The structure of inductive system on the algebras $\ca_n$
induces an analogous structure on the Hilbert spaces. In this way,
the inductive limit $\ch ^\infty $ will naturally carry a
representation of $\ca_\infty $. Moreover, $\ch ^\infty$ will
have a natural realization as the space of square integrable
functions $L^2(Q^\infty, \mu^\infty)$. All that is needed to carry this
construction out is to introduce a
suitable system of cyclic vectors $\{\chi_n\}$ one for each level.
A possible choice for $\chi_n$ is the
vector whose components are all equal to 1. Once the cyclic
vectors have been chosen, embeddings
$i_{{(n,m)}}:\ch ^n \rightarrow \ch ^{m}$, $n<m$, are defined by
\be
i_{{(n,m)}}(\hat a_n) \chi _n \equiv
\widehat{\Phi_{{(n,m)}}(a_n)}\chi_m,~~~a_n \in \ca_n~,~
\chi_n\in \ch^n~,~\chi_m\in \ch^m~, \label{embh}
\ee
where $\Phi _{{(n,m)}}$ are the embedding between the algebras $\ca_n$
given in (\ref{emb}).
The importance of having a cyclic vector defined at each
level is clear from this equation as this makes the embeddings
$i_{{(n,m)}}$ defined on all of $\ch ^n$.
With our choice of cyclic vectors these embeddings are isometries
\fn{The general choice of cyclic vectors which makes the embeddings
(\ref{embh}) isometric should satisfy the condition
$ | \chi^n_j|^2 = \sum_{l, \pi^{{(n,m)}}(q^n_j) = q^m_j} |\chi^m_l|^2
\mu_l^m$.}.
The inductive limit of the
$\ch^n$'s, namely $\ch ^{\infty }$, can be defined by means of
Cauchy sequences of vectors $\{\psi _n \}$ as was done for the
algebra $\Ainf$.  There exists natural isometric embeddings
$i_n:\ch ^n \rightarrow \ch ^\infty$ defined as
\be
i_n(\psi _n) \equiv [\{\phi_m\}] = \{i_{{(n,m)}}(\psi _n),~~m>n\}~,~~~~
\phi_n \in \ch^n~,   \label{embc}
\ee
where $[\cdot]$ denotes equivalence class.

Nevertheless, we find it convenient to present this inductive system
in terms of $L^2$ spaces of square-integrable functions. Think of the
$\ch ^n$ as $L^2(Q^n, \mu ^n)$.
The sequence of
measures $\mu ^n$ induces in turn a measure on the open sets of
$Q^\infty $, which can be extended, using standard methods \cite{KF},
to a
unique $\sigma $-additive absolutely continuous measure $\mu ^\infty $ on
$Q^\infty $. The embeddings $i_n$ in eq. (\ref{embc}), become the pull-backs of
$L^2(Q^n, \mu ^n)$ in $L^2(Q^\infty, \mu^\infty )$ of the projections $\pi _n$
{}from $Q^\infty$ to $Q^n$. Since the union of the $L^2(Q^n, \mu ^n)$ is dense
in
$L^2(Q^\infty, \mu ^\infty )$, and since $\ch ^\infty$ can be seen as the set
of equivalence classes of Cauchy sequences in $L^2(Q^{(\infty )},\mu
^{\infty})$, it follows that these two spaces can
be identified.

Finally, recall
again that $Q^{(\infty )}$ is a quasi fiber bundle on $M$. The
set containing all the fibers consisting of more that one point is
countable and thus has zero measure. It follows that
$L^2(Q^\infty, \mu^\infty)$ is naturally isometric to $L^2(M, \mu )$.

\sxn{Conclusions}

In this paper we have discussed some aspects of
the topology of a space $M$ when one considers lattice
discretizations of it. We have seen that the space being
approximated can be recovered from a projective limit of a
projective system associated with the lattice discretizations.

The projective limit is basically an
universal object (the Cantor set), and
the information of the original topology can be encoded in a
continuous projection from the Cantor set.

We have also discussed (in the spirit of Noncommutative Geometry) the
dual algebra of continuous functions for the lattices at the finite level,
and for the continuum limit. It is possible to recover the
algebra ${\cal C}(M)$ of continuous functions
of the space $M$ as a subalgebra of the continuous functions on the
Cantor set. Therefore even from the algebraic point of view the continuum
limit of lattices is an universal object, and the information over the
original starting point all lies in the choice of this subalgebra.

All of these aspects will be dealt with again in the context of
noncommutative lattice in \cite{comp}, where
the universality of the limits will be lost and both the space $M$ and
the algebra ${\cal C}(M)$ will arise naturally from the projective
and direct system, respectively.

\bigskip

\bigskip

\noindent
{\large \bf Acknowledgements }

We would like to thank G. Marmo and P. Michor for inviting us at ESI
and all people at the Institute for the warm and friendly atmosphere.
We thank A.P. Balachandran for many fruitful discussions and useful
advice. We also thank Mauro Carfora and Jerry Jungman for useful
discussions.

We thank the `Istituto Italiano per gli Studi Filosofici' in Napoli for
partial support.
The work of P.T-S. was also supported by the Department of Energy,
U.S.A. under
contract number DE-FG-02-84ER40173. The work of G.L. was
partially supported by the Italian `Ministero dell' Universit\`a e
della Ricerca Scientifica'.

\vfill\eject

\end{document}